\documentclass[12pt]{article}
\usepackage[dvips]{graphicx}
\usepackage{amsfonts}
\usepackage{amsmath}
\usepackage{amssymb}
\usepackage{latexsym}
\begin{document}

\title{Information entropy and nucleon correlations in nuclei}

\author{ S.E. Massen$^1$, V.P. Psonis$^1$, A.N. Antonov$^2$,\\
1) Department of Theoretical Physics,\\
 Aristotle University of Thessaloniki,
GR-54124 Thessaloniki, Greece,\\
2) Institute of Nuclear Research and Nuclear Energy,\\
Bulgarian Academy of Sciences, Sofia 1784, Bulgaria\\
 }
%\date{}
\maketitle

\vspace*{-1.5cm}

\begin{abstract}
The information entropies in coordinate and momentum spaces and
their sum ($S_r$, $S_k$, $S$) are evaluated for many nuclei using
"experimental" densities or/and momentum distributions. The
results are compared with the harmonic oscillator model and with
the short-range correlated distributions. It is found that $S_r$
depends strongly on $\ln A$ and does not depend very much on the
model. The behaviour of $S_k$ is opposite. The various cases we
consider can be classified  according to either the quantity of
the experimental data we use or by the values of $S$, i.e., the
increase of the quality of the density and of the momentum
distributions leads to an increase of the values of $S$. In all
cases, apart from the linear relation $S=a+b\ln A$, the linear
relation $S=a_V+b_V \ln V$ also holds. V is the mean volume of the
nucleus. If $S$ is considered as an ensemble entropy, a relation
between $A$ or $V$ and the ensemble volume can be found. Finally,
comparing different electron scattering experiments for the same
nucleus, it is found that the larger the momentum transfer ranges,
the larger  the information entropy is. It is concluded that $S$
could be used to compare different experiments for the same
nucleus and to choose the most reliable one.\\


\end{abstract}

%\newpage
%%%%%%%%%%%%%%%%%%%%%%%%%%%%%%%%%%%%%%%%%%%%%%%%%%%%%%%%
\section{Introduction\label{sec-1}}
%%%%%%%%%%%%%%%%%%%%%%%%%%%%%%%%%%%%%%%%%%%%%%%%%%%%%%%%%

Information theoretical methods are starting to be important tools
for studies of quantum mechanical systems. An example is the
application of the Maximum Entropy Principle \cite{Kapur89} (MEP)
to the calculation of the wave function in a potential
\cite{Canosa899092} using as constraints expectation values of
simple observables and reconstructing a quantum wave function from
a limited set of expectation values. The idea behind the MEP is to
choose the least biased result, compatible with the constraints of
the problem. Thus the MEP provides the least biased description
consistent with the available relevant information. This is done
by employing a suitably defined information entropy (IE) that
measures the lack of information associated with the distribution
of a quantum state over a given known basis. A measure of the IE
is Shannon's information entropy $S$ \cite{Shannon48}. For a
continuous probability distribution $p(x)$ ($\int p(x) d x=1$) $S$
is defined
\[
S=-\int p(x) \ln p(x) d x
\]

Shannon's IE has played an important role in the study of quantum
mechanical systems, in clarifying fundamental concepts of quantum
mechanics and in the synthesis of probability densities in
position and momentum space
 \cite{Bialy75,Gadre848587,Ohya93,Nagy96,Majer96,Panos97,%
Lalazi98,Massen9801,Massen02,Panos01b,Massen03,Moustakidis04,Shi04}.
An important step was the discovery of an entropic uncertainty
relation \cite{Bialy75}. For a three-dimensional system it has the
form
\begin{equation}
S= S_r + S_k \ge 3 (1+ \ln \pi) \simeq 6.434 \quad (k=p/\hbar),
 \label{S1}
\end{equation}
where
\[
S_{r}=-{\int}{\rho}({\bf r}) \ln {\rho}( {\bf r})d {\bf r},\quad
S_{k}=-{\int}n( {\bf k})\ln n({\bf k})d {\bf k}
 %\label{Sr-k-1}
\]
are Shannon's IEs in coordinate and momentum space and $\rho ({\bf
r})$, $n({\bf k})$ are the density distribution (DD) and momentum
distribution (MD), respectively, normalized to 1.

Inequality  (\ref{S1}) is an information-theoretical uncertainty
relation stronger than Heisenberg's \cite{Bialy75} and does not
depend on the unit of length in measuring $\rho({\bf r})$ and
$n({\bf k})$, i.e. the sum $S=S_r+S_k$ is invariant to uniform
scaling of coordinates, while the individual entropies $S_r$ and
$S_k$ are not.  The physical meaning of $S$ is that it is a
measure of quantum-mechanical uncertainty and represents the
information content of a probability distribution, in our case of
the nuclear density and momentum distributions. Inequality
(\ref{S1}) provides a lower bound for $S$ which is attained for
Gaussian wave functions.

We quote March who refers to the information entropy with the
following words: ``Further work is called for before the
importance of $S_r$ and $S_k$ in atomic theory can be
assessed''.\cite{March92,Ziesche95} We could extend that statement
for fermionic and correlated bosonic systems as well.

Shannons's IEs $S_r$ and $S_k$ have  been recently studied for the
densities of various systems \cite{Panos97,Massen9801,Massen02}:
the nucleon DD of nuclei, the valence electron DD of metallic
clusters and the DD of correlated Bose alkali atoms. It has been
found that the same functional form $S=a+b \ln A$ for the entropy
sum as function of the number of particles $A$ holds approximately
for the above systems in agreement with Ref. \cite{Gadre848587}
for atomic systems. Another interesting result \cite{Hall94} is
the fact the entropy of an $N-$phonon state subjected to Gaussian
noise increases linearly with the logarithm of $N$. In Refs.
\cite{Massen03,Moustakidis04} the dependence of $S$ on the
short-range correlations (SRC) parameter of the nucleons in
nuclei, and of the particle interaction in various uniform Fermi
systems (nuclear matter, $^3$He liquid, and electron gas) has been
found. This dependence as well as the linear dependence of $S$ on
$\ln A$ were used in Ref. \cite{Massen03} to determine the SRC
parameter of nucleons in $s-$, $p-$, and $sd-$shell nuclei.
In Ref. \cite{Lalazi98} another definition of IE according to
phase-space considerations \cite{Ghosh84} was used and an
information theoretical criterion for the quality of a nuclear DD
was derived, i.e. the larger $S$ the better the quality of the
nuclear model. In Ref. \cite{Moustakidis01} the DD, the MD and the
Shannon's IEs were calculated for nuclei using three different
cluster expansions. The parameters of the various expressions were
determined by least-squares fit of the theoretical charge form
factors to the experimental ones. It was found that the larger the
entropy sum, the smaller is the value of $\chi^2$, indicating that
the maximal $S$ is a criterion of the quality of a given nuclear
model according to the MEP. Only two exceptions to that rule were
found out of many cases examined.
 Before proceeding, it is appropriate to mention
that additional applications of entropy have attracted interest in
recent years \cite{BAB95,BAB98}, but in a different spirit, in
nuclear physics problems, such as in analysis of shell-model
eigenvectors. The authors in Ref. \cite{BAB98} defined a
correlational entropy. We note, however, that this is a von
Neumann entropy, which they applied in the framework of the
nuclear shell model. In our case we use the definition of the IE
according to Shannon applied to the density distribution in
coordinate and in momentum space of a nuclear system. Finally,
alternative measures of the IE have been proposed by Onisescu
\cite{Onisescu66} and by Brukner and Zeilinger \cite{Brukner01}.

The motivation of the present work is to extend our previous study
of IE in nuclei using as many experimental data available in the
literature as possible. In this way, it will be possible to study
important features of realistic nuclear systems, e.g. the effects
of nucleon-nucleon correlations on basic nuclear characteristics,
such as density and momentum distributions. Thus, instead of
starting our study from a given nuclear model we start from the
"experimental"  DDs from Refs. \cite{DeVries87} and \cite{Burov98}
(obtained from electron scattering by nuclei and muonic atoms)
or/and from estimations of "experimental" MDs from Refs.
\cite{Antonov04,Antonov05} based on superscaling analysis
\cite{Donnelly99} of inclusive electron scattering from nuclei as
well as on the coherence density fluctuation model (CDFM)
\cite{Antonov79,Antonov85,Antonov88,Antonov899486} which helps to
calculate the MD in connection with the DD and vice versa. This
study helps in two ways:

First, it helps to strengthen the empirical property $S=a+b\ln A$
which has been proposed in previous studies for various fermion
systems \cite{Gadre848587,Panos97,Massen9801,Massen02}. Also it
enables us to strengthen the conclusion of Ref. \cite{Lalazi98}
that the more experimental data we use in the theory the larger is
the IE of a nucleus, i.e. the larger $S$ the better the quality of
the DD of a nucleus is. For the five models which we applied to
various nuclei from $^4$He to $^{238}$U it was found that the same
functional form of $S$ holds, while the values of the parameters
$a$ and $b$ depend on the model. The harmonic oscillator (HO)
shell-model is closer to the lower limit of inequality (\ref{S1})
which is attained only for the $^4$He nucleus. The results for
more realistic models deviate from the HO ones. The deviation
becomes larger when more experimental data are included in the
model. The various cases we considered can be classified according
to either the quantity of the experimental data we use or the
values of $S$, i.e., the increase of the quality of the density
and of the momentum distributions leads to an increase of the
values of $S$. As the DD and the MD based on experimental data
should be considered as the least biased ones, our results are in
accordance with the MEP. Another characteristic result is that the
various lines $S_{model}=a_{model} + b_{model} \ln A$ are almost
parallel, i.e., there is a kind of scaling of the values of $S$
for the various models.

Second, our work helps to connect the IE with other physical
quantities such as the root-mean-square (RMS) radius of the
nucleus or the mean volume in which the nucleons are confined in a
nucleus. If Shannon's IE is considered as ensemble entropy, this
connection could be used to relate the mean volume of the nucleus
or the mass number with the "ensemble volume" through the relation
$V_{ensemble}=e^S$ \cite{Hall99}. This ensemble volume provides a
direct measure of uncertainty of the system which is advantageous
when one wishes to compare the spread of two ensembles of a given
type. Finally, the present work might also help to choose the most
reliable experiment when more than one experiments are made for
the same nucleus.

The paper is organized as follows: In Sec. II we briefly review
the theory of the CDFM and describe the way in which the IEs
$S_r$, $S_k$ and their sum can be calculated from the DD or/and
the MD. In Sec. III the numerical calculations of IEs for various
nuclei, using three different approaches, are presented and
compared with previous calculations. The possibility of choosing
the most reliable experiment, when more than one experiments are
made for the same nucleus, is also discussed. In Sec. IV, a
dependence of the IE sum on the mean volume of the nucleus is
proposed and a connection between the ensemble volume of the
nucleus and the mean volume or the mass number is given. Finally,
Sec. V contains our conclusions.

\section{The information entropy from a given distribution}

In order to find the information entropy of various nuclei, the
distributions of the density and the momentum should be known.
These distributions can be found using various models. Another way
is to employ given DDs, i.e., the phenomenological ones given e.g.
from Refs. \cite{DeVries87,Burov98} or the "experimental" MDs
using the superscaling analysis in nuclei
\cite{Donnelly99,Antonov04,Antonov05}. The idea is to use the CDFM
to find the MD in connection with the DD or to find the DD in
relation to the MD of a nucleus.

The CDFM \cite{Antonov79,Antonov85,Antonov88,Antonov899486} is
related to the $\delta-$function limit of the generator coordinate
method \cite{Griffin57} in which the $A-$body wave function of a
nucleus is written in the form
\begin{equation}
\Psi({\bf r}_1,{\bf r}_2,\cdots,{\bf r}_A)= \int
F(x_1,x_2,\cdots)\Phi({\bf r}_1,{\bf r}_2,\cdots,{\bf r}_A;
x_1,x_2,\cdots)d x_1 d x_2\cdots\ ,
 \label{psi-1}
 \end{equation}
where the generating function $\Phi(\{{\bf r}_i\};x_1,x_2,\cdots)$
depends on the coordinates of the nucleons (radius-vector, spin,
isospin) and on the generator coordinates  $x_1,x_2,\cdots$.
$\Phi$ is usually chosen to be a Slater determinant built up from
single-particle wave functions corresponding to a given
construction potential parameterized by the generator coordinates.
The weight function $F(x_1,x_2,\cdots)$ can be determined (using
the variational principle) as a solution of the Hill-Wheeler
integral equation
\begin{equation}
\int [ {\cal H}(x,x') - E {\cal I}(x,x')] F(x')\, d x'=0,
 \label{Hill}
 \end{equation}
where ${\cal H} (x,x')=\langle \Phi (\{ {\bf r}_i\},x)|{\hat H}|
\Phi (\{ {\bf r}_i\},x')\rangle$ and ${\cal I} (x,x')=\langle \Phi
(\{ {\bf r}_i\},x)| \Phi (\{ {\bf r}_i\},x')\rangle$,
$i=1,2,\cdots,A$ and $x$ denotes a set of $x_1,x_2,\cdots$.

In the CDFM the DD and the MD are expressed by means of the same
weight function $F(x)$
\begin{equation}
 \rho({\bf r}) = \int_0^{\infty} \frac{3A}{4\pi x^3}\, |F(x)|^2
 \Theta (x- |{\bf r}|)\, d x
 \label{den-r-1}
 \end{equation}
 and
\begin{equation}
n({\bf k}) =\frac{4}{(2\pi)^3}  \int_0^{\infty} \frac{4\pi
x^3}{3}\, |F(x)|^2  \Theta (k_{F}(x)- |{\bf k}|)\, d x
 \label{den-k-1}
 \end{equation}
both normalized to the mass number $A$
\begin{equation}
 \int \rho({\bf r})\, d {\bf r}=A,\qquad \int n({\bf k})\, d {\bf
 k}=A.
  \label{den-rk-A}
 \end{equation}
$\Theta$ is the unit step function, $x$ is the generator
coordinate and $k_F(x)$ is the Fermi momentum of a piece of
nuclear matter with radius $R=x$
\begin{equation}
k_F(x)=\left(\frac{3\pi^2}{2}\rho_0(x)\right)^{1/3}=\frac{\alpha}{x},\qquad
\alpha=\left(\frac{9\pi A}{8}\right)^{1/3}.
 \label{k-Fermi}
 \end{equation}
For DD normalized to $A$ the weight function obeys the constraint
\begin{equation}
 \int_0^{\infty} |F(x)|^2 d x=1.
\end{equation}

Various paths can be followed to find the function $F(x)$. Here we
follow the approach proposed in Refs. \cite{Antonov79,
Antonov85,Antonov88} and used also in \cite{Antonov04,Antonov05}.
As can be seen from Eq. (\ref{den-r-1}), for known DD the weight
function $F(x)$ can be determined by
 \begin{equation}
|F(x)|^2=-\frac{1}{\rho_0(x)} \left. \frac{d \rho}{d
r}\right|_{r=x},
 \label{Fx-1}
\end{equation}
where $\rho_0(x)=\frac{3 A}{4\pi x^3}$ and $\rho(r)$ satisfies the
constraint $\frac{d \rho}{d r} \le 0$.

Substituting $|F(x)|^2$ from  Eq. (\ref{Fx-1}) in the right-hand
side of  Eq. (\ref{den-k-1}), $n(k)$ takes the form
\begin{equation}
n(k)=\frac{8}{9\pi A} \left[ 6\int_0^{\alpha/k} r^5\rho(r)\, d r -
\left(\frac{\alpha}{k}\right)^6
\rho\left(\frac{\alpha}{k}\right)\right].
 \label{den-k-2}
 \end{equation}
Eq.  (\ref{den-k-2}) shows the MD as a functional of the density
distribution. This point is discussed in Refs.
\cite{Antonov88,Antonov86} within the framework of the density
functional theory.

Thus, within the CDFM the MD of a nucleus can be found from Eq.
(\ref{den-k-2}). From $\rho(r)$, and $n(k)$ the information
entropies $S_r$, $S_k$ defined by the relations
\begin{equation}
S_{r}=-{\int}{\rho}({\bf r}) \ln {\rho}( {\bf r})d {\bf r},\quad
S_{k}=-{\int}n( {\bf k})\ln n({\bf k})d {\bf k},
 \label{Sr-k-1}
\end{equation}
and their  sum $S=S_r+S_k$ can be calculated. We note that for the
calculation of $S_r$ and $S_k$ we use DD and MD normalized to 1.

In the recent paper \cite{Antonov05} the momentum distribution of
nuclei $^4$He, $^{12}$C, $^{27}$Al, $^{56}$Fe, and $^{197}$Au was
calculated using the analysis of the superscaling phenomenon in
inclusive electron scattering from nuclei \cite{Donnelly99}. From
those distributions and from Eq. (\ref{den-k-1}) the weight
function $F(x)$ can be calculated as
\begin{equation}
 |F(x)|^2=-\frac{1}{n_0(x)} \left. \frac{d n}{d
k} \right|_{k=x},
 \label{Fx-2}
\end{equation}
where $n_0(x)=\frac{3 A}{4\pi x^3}$ and $n(k)$ satisfies the
constraint $\frac{d n}{d k} \le 0$.

Substituting $|F(x)|^2$ from  Eq. (\ref{Fx-2}) in the right-hand
side of  Eq. (\ref{den-r-1}), the DD can be expressed by
\begin{equation}
\rho(r)=\frac{8}{9\pi A} \left[ 6\int_0^{\alpha/r} k^5 n(k)\, d k
- \left(\frac{\alpha}{r}\right)^6
n\left(\frac{\alpha}{r}\right)\right].
 \label{den-r-2}
 \end{equation}

Thus, within the CDFM the DD of a nucleus can be estimated
approximately by means of the MD. Using $\rho(r)$ and $n(k)$ the
entropies $S_r$, $S_k$, and $S$ can be calculated from Eqs.
(\ref{Sr-k-1}).

We note the symmetry of the expressions (\ref{den-k-2}) and
(\ref{den-r-2}) for both equivalent basic characteristics, the
nucleon momentum and local density distributions.

\section{Numerical results and discussion}

For the calculation of the information entropy of a nucleus we
used three different approaches. In the first one we used the
"experimental" DDs for various nuclei from $^4$He to $^{238}$U
from Refs. \cite{DeVries87,Burov98}.  For the various DDs existing
in the literature we used only two or three parameter Fermi (2pF,
3pF) distributions from \cite{DeVries87}
\begin{equation}
\rho(r)=\rho_0\frac{1+wr^2/c^2}{1+\exp[(r-c)/\alpha]}
 \label{23pF}
\end{equation}
($w=0$ in the 2pF distributions), and the symmetrized Fermi
distributions from \cite{Burov98}
\begin{equation}
\rho(r)=\rho_0\frac{\sinh (c/\alpha)}{\cosh(r/\alpha) +
\cosh(c/\alpha) } .
 \label{SymF}
\end{equation}
The reason we avoided the use of other phenomenological
distributions is that there usually exist oscillations in the
central region of the densities of the nuclei which destroy the
constraint $d\rho /dr\le 0$. This is not the case for the
Fermi-type distributions.

From those distributions and from Eq. (\ref{den-k-2}) the MD for
various nuclei can be found. In the second approach we used the
"experimental" MD from the superscaling analysis of Ref.
\cite{Antonov05} for the nuclei $^4$He, $^{12}$C, $^{27}$Al,
$^{56}$Fe, and $^{197}$Au. As shown in \cite{Antonov05}, in the
CDFM the MD is approximately related to the $\psi'-$scaling
function $f(\psi')$ (introduced in \cite{Donnelly99}) by
 \begin{equation}
 n(k)=-\frac{1}{3\pi k^2 k_F} \left.  \frac{\partial
 f(\psi')}{\partial(|\psi'|)}\right|_{|\psi'|=k/k_F}
 \label{k-k-psi}
 \end{equation}
where $k_F$ is the Fermi momentum which can be calculated within
the  CDFM \cite{Antonov04,Antonov05}. Using the experimental data
for $f(\psi')$ obtained from inclusive electron scattering from
nuclei \cite{Donnelly99} we estimated the MD $n(k)$.  From those
MDs, using Eq. (\ref{den-r-2}) the DD of these nuclei were found.
In the third approach we used the "experimental" DDs, for the
above mentioned five nuclei, as in the first approach and for the
MDs the "experimental" values from the superscaling analysis as in
the second approach.

\begin{figure}[h]
\begin{center}
\hspace*{-1cm}\begin{tabular}{ccc}
 {\includegraphics[width=5.5cm]{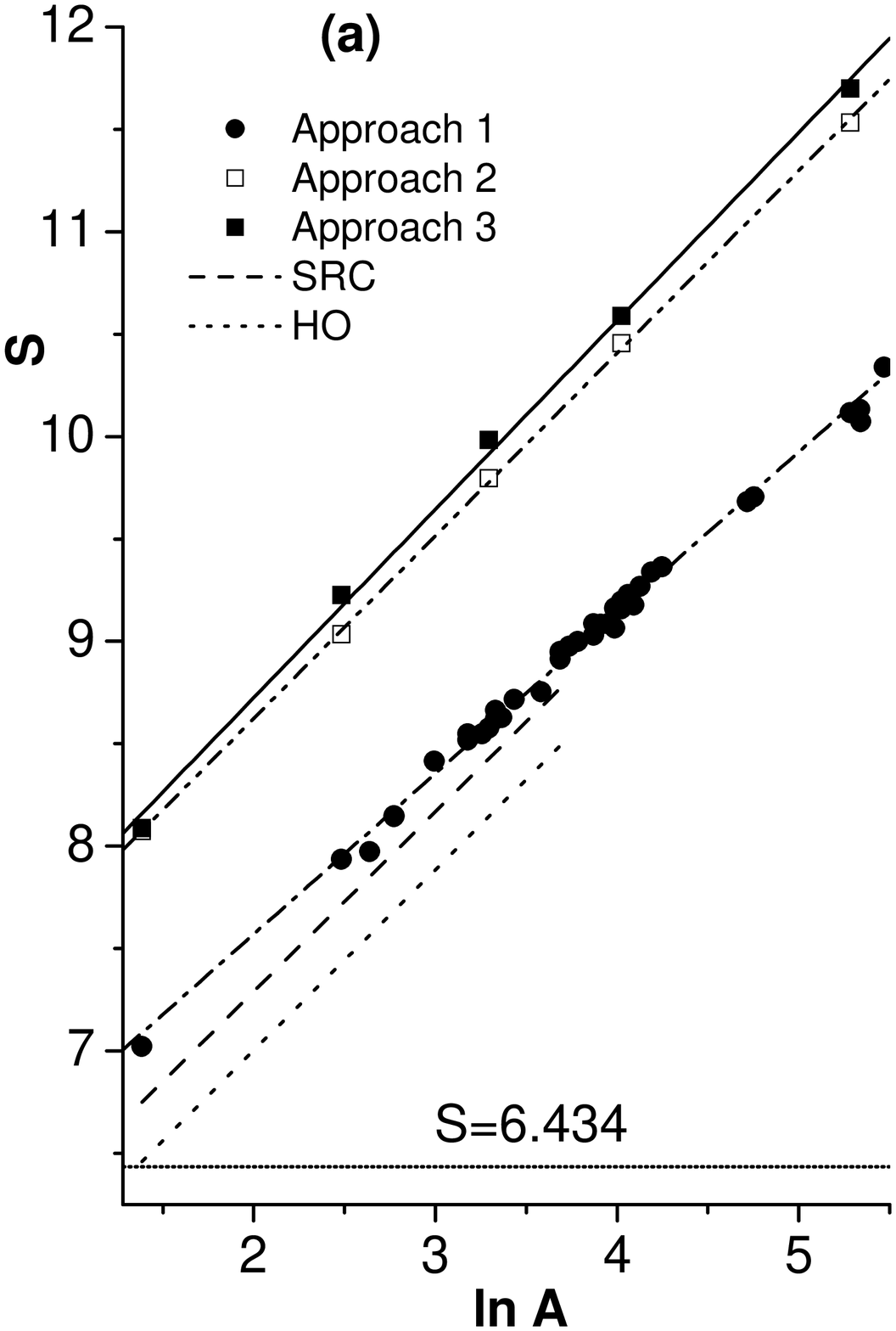}}&
 {\includegraphics[width=5.5cm]{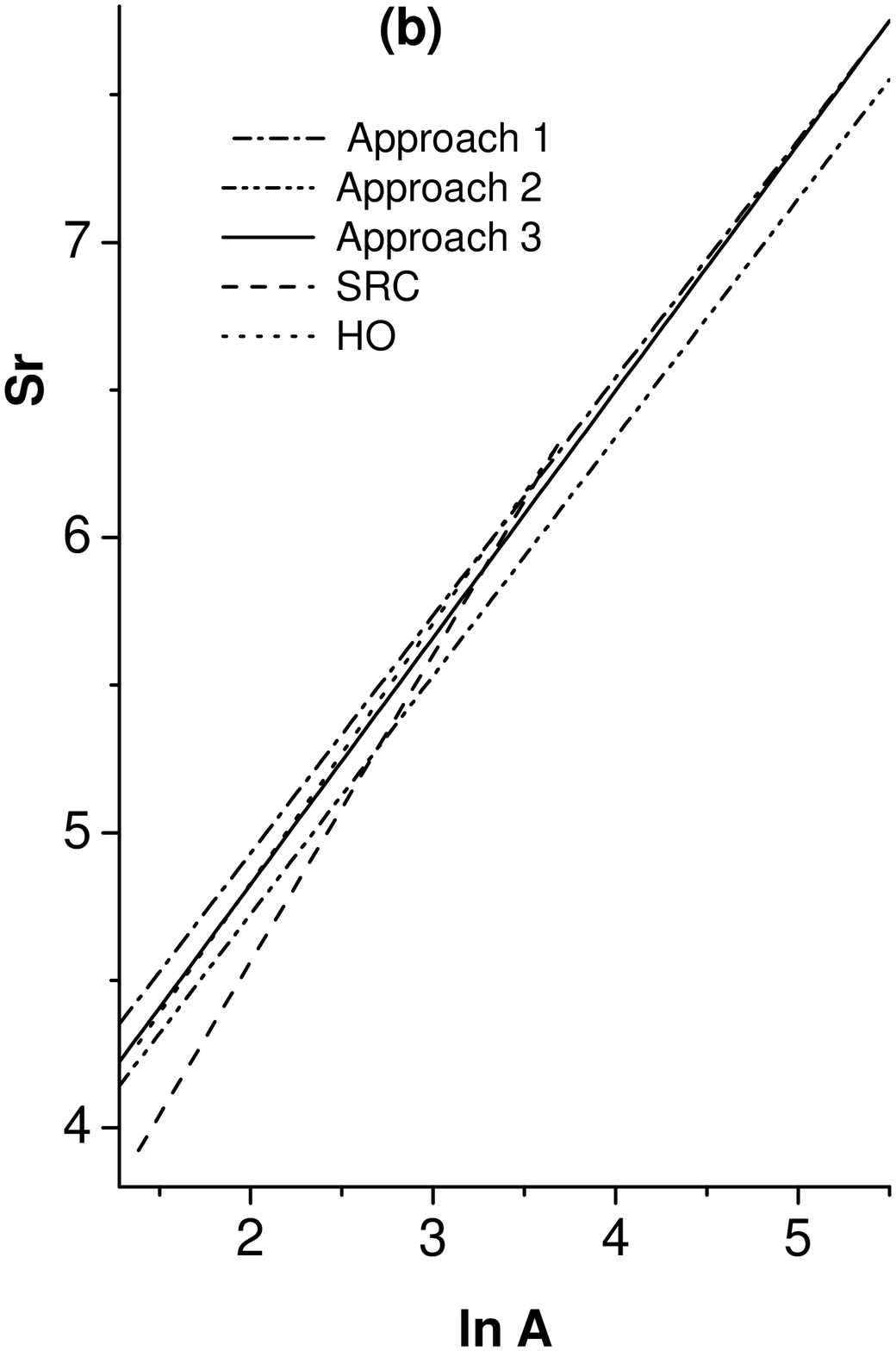}}&
 {\includegraphics[width=5.5cm]{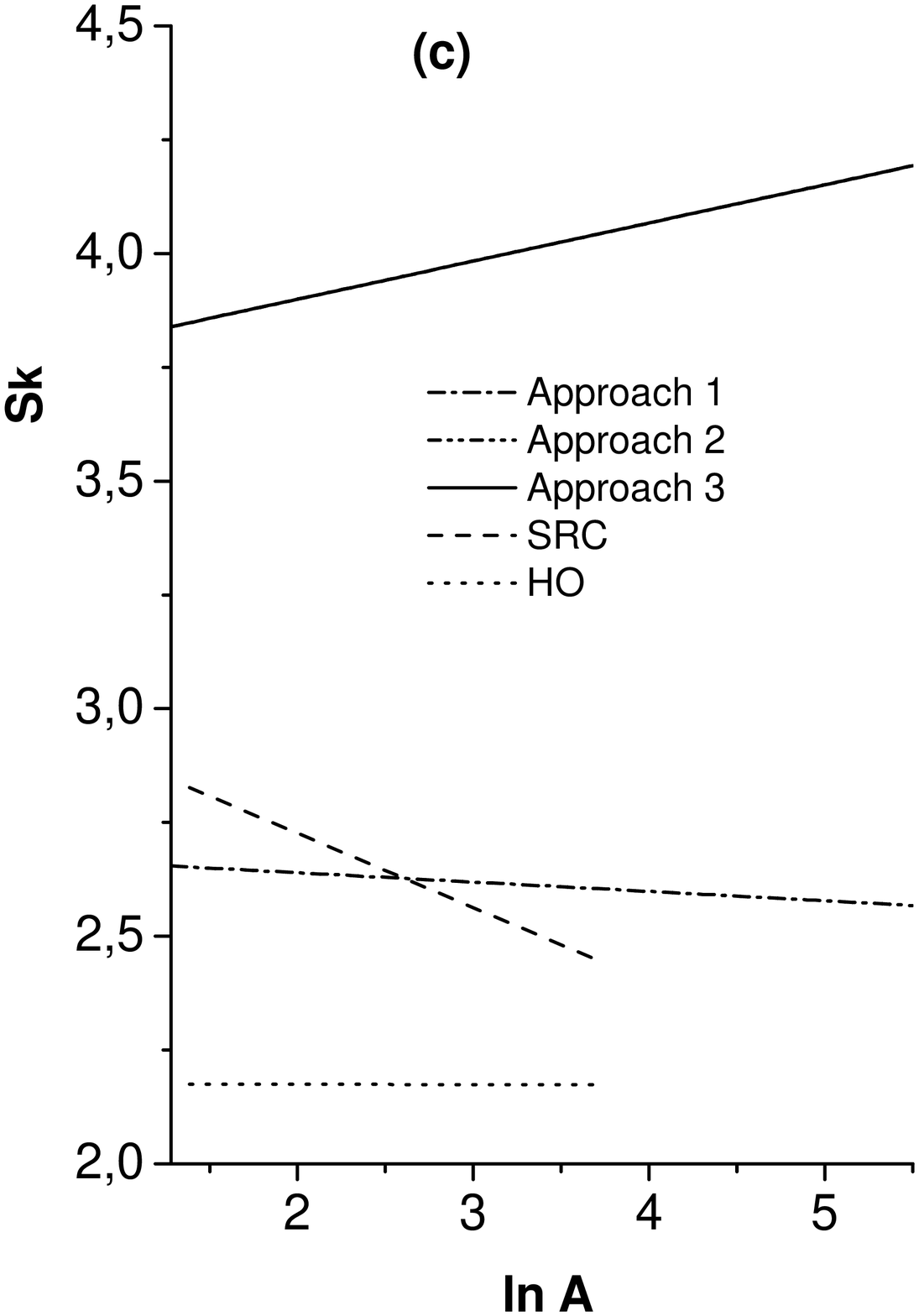}}
\end{tabular}
\end{center}
\vspace*{-1cm}
 \caption{The information entropies in nats: (a)  $S=S_r+S_k$,
(b) $S_r$, and (c) $S_k$ for various nuclei versus the logarithm
of the mass number $A$. The lines correspond to the fitting
expressions $S=a+b\ln A$ and $S_{r,k}=a_{rk}+b_{r,k}\ln A$ ,
respectively.  For the various cases see text. The limiting line
corresponding to the lower bound $S=6.434$ is also shown.}
\end{figure}

The evaluated values of $S=S_r +S_k$ in the three approaches are
shown by points in Fig. 1a. One can see that $S$ obeys the
universal property
\begin{equation}
  S=S_r +S_k =a + b \ln A .
  \label{S-ab}
\end{equation}
The same holds for the IEs $S_r$ and  $S_k$ which obey the
relation
\begin{equation}
 S_{r,k}=a_{r,k} + b_{r,k} \ln A
  \label{S-ab-rk}
\end{equation}

The parameters $a$, $b$ and $a_{r,k}$, $b_{r,k}$ were determined
by least squares fit of the values of $S$  and $S_r$, $S_k$
calculated from Eqs. (\ref{S-ab}) and (\ref{S-ab-rk}),
respectively, to the corresponding evaluated values of the IEs
from Eqs. (\ref{Sr-k-1}). Their values found in the three
approaches are shown in Table I. In Figs. 1b and 1c, where the
lines $S_r(A)$ and $S_k(A)$ (Eq. (\ref{S-ab-rk})) are shown, we
have not displayed the calculated values of $S_r$ and $S_k$, from
Eqs. (\ref{Sr-k-1}) as the values of the various points are very
close to each other in many cases.
\begin{table}
\caption{The values of the parameters $a$, $b$, $a_{r,k}$,
$b_{r,k}$, $a_V$, and $b_V$ of relations (\ref{S-ab}),
(\ref{S-ab-rk}), and (\ref{S-V}) in the various cases.}
 \begin{center}
\begin{tabular}{l cc cc cc cc}
\hline\\
Case
%&\multicolumn{2}{c}{$S_r$}&\multicolumn{2}{c}{$S_k$}&\multicolumn{2}{c}{$S$}\\
 & $a$ &$b$ & $a_r$ &$b_r$ & $a_k$ &$b_k$&$a_V$&$b_V$\\
  \hline\\
 HO        & 5.2391& 0.8816& 3.0633& 0.8822& 2.1758&-0.0006&3.4783&0.9472 \\
 SRC       & 5.5330& 0.8778& 2.4807& 1.0409& 3.0524&-0.1631&4.2723&0.8512 \\
Approach 1 & 6.0011& 0.7847& 3.3205& 0.8053& 2.6807&-0.0205&4.1166&0.9272\\
Approach 2 & 6.8396& 0.8919& 3.1078& 0.8080& 3.7318& 0.0836&2.0255&1.3969 \\
Approach 3 & 6.8845& 0.9201& 3.1527& 0.8362& 3.7318& 0.0839&4.8413&1.0635 \\
\hline
\end{tabular}
\end{center}
\end{table}

For completeness we also give in Table I the corresponding values
of those parameters for the HO case, and for the case of
short-range correlated DD and MD which were found with three
different expansions of the one-body density matrix in Ref.
\cite{Moustakidis01}. We mention that the values of $S_r$, $S_k$,
and $S$ were found in \cite{Moustakidis01} for the $s-$, $p-$, and
$sd-$shell nuclei for DD and MD normalized to $A$. From those
values of $S_r$, $S_k$ and $S$ using the relation
\cite{Massen9801}
\begin{equation}
S_{r,k}[norm=1] = \frac{1}{A} S_{r,k}[norm=A] + \ln A
\end{equation}
the corresponding IEs for DD and MD normalized to 1 were found.
The values of the IEs $S_r$, $S_k$ and $S$ for the three
approaches as well as for the HO and the SRC approaches calculated
using the values of the parameters $a$ and $b$ of Table I are also
shown in Fig. 1 by the corresponding lines.

The various cases we examined can be placed in order either by the
quantity of the experimental data that were used or by the values
of the information entropy obtained for the various nuclei. This
is a consequence of Fig. 1a and also of the following discussion.

In the HO model there is only one free parameter which can be
determined either by the experimental RMS charge radius or by fit
of the theoretical charge form factor ($F_{ch}(q)$) to the
experimental one. The results of the HO case presented in Fig. 1
were obtained in Ref. \cite{Moustakidis01} by fit of the
theoretical $F_{ch}(q)$ to the experimental data. In this case
only the low momentum transfer data are reproduced. If we include
SRC as in Ref. \cite{Moustakidis01} there are two free parameters
which are determined by fit of the theoretical $F_{ch}(q)$ to the
experimental one. In this case more diffraction minima are
reproduced than in the HO case. This is reflected in the values of
$S$. The inequality  $S_{SRC} > S_{HO}$ holds for all the $s-$,
$p-$, and $sd-$shell nuclei we have examined.

In the first approach of the present work the Fermi-type
distributions, which are employed, are phenomenological
distributions reproducing better the electron scattering
experiments than in the previous two cases. This is reflected in
the values of $S$. For all nuclei from $^4$He to $^{238}$U we have
examined, the inequality $S_{approach1}>S_{SRC}$ holds.

In the second approach the "experimental" data of the MD
\cite{Antonov05} were used. In this case the high-momentum
components of the MD were included in the calculations in a more
reliable way than in the first approach. This results in an
increased contribution of $S_k$ to the information entropy sum.
Thus, the inequality $S_{approach2}>S_{approach1}$ holds for the
five nuclei we have examined.

Finally, in the  third approach information from experimental data
both from DD \cite{DeVries87,Burov98} and from MD \cite{Antonov05}
were included. This leads to increased values of $S$ in comparison
with the corresponding values in the  second approach where only
information from the "experimental" MD were taken into account in
the calculation of $S$.

Thus, we should conclude that the increase of the quality of the
DD or/and of the MD leads to an increase of the values of the
information entropy sum. As the DD and the MD based on
experimental data should be considered as the least biased ones,
the previous statement is in accordance with the MEP. Another
characteristic feature of $S_{model}$ can be seen from Fig. 1a,
i.e. the lines $S_{model}=a_{model} + b_{model} \ln A$ are almost
parallel, i.e. there is a kind of scaling of the values of $S$ for
the various cases we have examined.

From Figs. 1b and 1c and from the values of the parameters $b_r$
and $b_k$ it can also be concluded that $S_r$ depends strongly on
$\ln A$ while $S_k$ does not. Thus, the linear dependence of $S$
on $\ln A$ is mainly due to the IE in coordinate space. The strong
dependence of $S_r$ on $\ln A$ should be related to the volume of
the nucleus where the nucleons are confined. The weak dependence
of $S_k$ on $\ln A$ is related to the fact that the high-momentum
components of the MD are independent of the mass number of the
nucleus. This is a well known fact (see, e.g.
\cite{Antonov88,Antonov04,Antonov05}).

Another feature of Fig. 1b is that $S_r$ does not depend strongly
on the model which is used. The relative difference of the values
of $S_r$ obtained in the various cases is about 10\% or less.
$S_k$ (Fig. 1c) depends strongly on the model which is employed.
It becomes larger when the model includes high-momentum components
of the MD which are related to the presence of SRC. The relative
difference of $S_k$ for the various cases we have examined is
about 50\%.

%%%%%%%%%%%%%%%
%%%%%%%%%%%%%%%

From Fig. 1a (see also Fig. 2 in Sec. IV) one can see that the
points corresponding to five nuclei in approaches 2 and 3 lie
almost on the lines $S=a+b\ln A$. This is not the case in approach
1. In this approach while most of the points are on the line
$S=a+b\ln A$, there are few of them (e.g. the points corresponding
to nuclei $^{14}N$, $^{27}Al$ and $^{209}$Bi) which are relatively
far from that line. For these nuclei the distances of the
evaluated values of $S$ from the line $S=a+b\ln A$ are within the
errors of the parameters of the corresponding Fermi distributions.
It is also mentioned that for $^{209}$Bi the 2pF distribution
reproduces only the low momenta transfer of the electron
scattering in the $q-$range $=0.07 - 0.53\ {\rm  fm}^{-1}$.
\cite{DeVries87}

In Ref. \cite{DeVries87} there are cases where various 2pF and 3pF
distributions reproduce different experimental data for the same
nucleus. The deviations of some points from the line $S=a+b\ln A$
in approach 1 lead us to examine these distributions by comparing
the evaluated values of $S$ for the same nucleus. In Table II we
give the calculated values of $S$ for  various nuclei with Fermi
distributions from the analysis of different experiments and the
corresponding ranges of the momentum transfer.
\begin{table}[h]
\caption{The values of the information entropy $S$ for various
nuclei in approach 1. The calculations were made with the
phenomenological 2pF or/and 3pF distributions of Ref.
\cite{DeVries87}.}
\begin{center}
\begin{tabular}{l c c|l c c}
\hline
%&  &  &  & &  \\
Nucleus & $S$ & $q-$range & Nucleus & $S$ & $q-$range \\
&  [nats]& [fm$^{-1}$] &  &[nats] & [fm$^{-1}$] \\
\hline
$^{19}$F  & 8.3890 &0.55-1.01& $^{64}$Zn&9.2948& 0.30-1.09\\
$^{19}$F  & 8.3947 &0.46-1.79& $^{64}$Zn&9.2603&0.15-0.79\\
&  &  &  & &  \\
$^{20}$Ne & 8.3977 &0.22-1.04& $^{66}$Zn&9.3376&0.96-1.63\\
$^{20}$Ne & 8.4137 &0.21-1.12& $^{66}$Zn&9.2881&0.15-0.79\\
$^{20}$Ne & 8.4262 &0.49-1.80& \\
          &        &         & $^{68}$Zn&9.3253&0.96-1.63\\
$^{24}$Mg & 8.5162 &0.58-1.99& $^{68}$Zn&9.2920&0.15-0.79\\
$^{24}$Mg & 8.4660 &0.74-3.46& & & \\
$^{24}$Mg & 8.5175 &0.20-1.15& $^{70}$Zn&9.3643&0.30-1.09\\
          &        &         & $^{70}$Zn&9.3245&0.15-0.79\\
$^{50}$Cr & 9.0200 &0.15-0.79& & & \\
$^{50}$Cr & 9.0829 &0.97-1.62&$^{142}$Nd&9.8978&0.55-2.97\\ \
          &        &         &$^{142}$Nd&9.8211&0.23-0.59\\
$^{52}$Cr & 9.0243 &0.15-0.79&$^{142}$Nd&9.8762&0.22-0.73\\
$^{52}$Cr & 9.0860 &0.97-1.62&          &      & \\
          &        &         &$^{146}$Nd&9.9897&0.55-2.97\\
$^{54}$Cr & 9.0951 &0.15-0.79&$^{146}$Nd&9.8871&0.22-0.73\\
$^{54}$Cr & 9.1623 &0.97-1.62&          &      & \\
          &        &         &$^{150}$Nd&10.0284&0.55-2.97\\
$^{54}$Fe & 9.0633 &0.15-0.79& $^{150}$Nd&9.9305&0.22-0.73\\
$^{54}$Fe & 9.1014 &0.97-1.62& $^{150}$Nd&9.8638&0.37-2.29\\
          &        &         &           &      & \\
$^{56}$Fe & 9.1113 &0.15-0.79& $^{238}$ U&10.2478&0.37-0.97\\
$^{56}$Fe & 9.1585 &0.97-1.62& $^{238}$ U&10.2267&0.46-2.08\\
 \hline
\end{tabular}
\end{center}
\end{table}
It is seen that in almost all the cases the larger $q-$range
corresponds to larger value of $S$ for the same nucleus. From the
many cases of Table II we found only three  exceptions which
correspond to nuclei $^{24}$Mg, $^{150}$Nd and $^{238}$U. The
disagreement of the results for these nuclei to the above rule is
due to the following reasons:

In $^{24}$Mg the 3pF distributions which have been used in the
cases of momentum transfer ranges: $q=0.58-1.99\ {\rm fm}^{-1}$
and $q=0.74-3.46\ {\rm fm}^{-1}$ give charge distributions which
become $0$ for relatively small values of the radius  ($r\approx
7\ {\rm fm}^{-1}$). This is not the case for the 2pF distribution
($q-$range$=0.20-1.15\ {\rm fm}^{-1}$) which becomes $0$ for much
larger values of $r$. The existence of the logarithm of the
density in the integral of $S_r$ makes the IE sensitive to the
tail of the DD. This disagreement could be removed if we used the
errors of the parameters of the 3pF distributions.

In the nucleus $^{150}$Nd the value of $S$ corresponding to the
$q-$range$=0.22-0.73$ fm$^{-1}$ is larger than the value of $S$
corresponding to the $q-$range$=0.37-2.29$ fm$^{-1}$. We expected
the inverse order. This disagreement should be due to the fact
that the real analysis of the electron scattering data was made
with a deformed Fermi distribution in the latter case (see the
corresponding remark of Ref. \cite{DeVries87}). That deformation
was not taken into account in our calculations.

In the case of the nucleus $^{238}$U the 2pF distributions do not
give the experimental charge RMS radius 5.84 fm and 5.854 fm
corresponding to the momentum transfer ranges: $q=0.37-0.97\ {\rm
fm}^{-1}$ and $q=0.46-2.08\ {\rm fm}^{-1}$, respectively. The
values, we found, are 5.731 fm and 5.712 fm, respectively. That
difference should come from the 2pF distributions we have used,
instead of using the deformed Fermi distributions of Refs.
\cite{Cooper76, Creswell77} (see also the corresponding remarks of
Ref. \cite{DeVries87}).

Finally, in $^{142}$Nd while the values of $S$ are increasing with
the $q-$range, the increase is quite small from $q-$range$=0.22 -
0.73\ {\rm fm}^{-1}$ to $q-$range$=0.55 - 2.97\ {\rm fm}^{-1}$.
The corresponding values of $S$ are 9.8762 and 9.8978,
respectively. The relatively small values of $S$ in the latter
case is due to the fact that the constraint $d \rho/d r \le 0$
does not hold for all the values of $r$. This has as a result the
high-momentum components of the MD not to be reproduced within the
CDFM as correctly as in the former case.

From the above discussion we should conclude that Shannon's
information entropy could be used to compare different experiments
for the same nucleus and to choose the most reliable one.

\section{The dependence of $S$ on the mean volume of the nucleus}

The strong dependence of $S_r$  and the nearly independence of
$S_k$ on $\ln A$ leads us to connect $S$ with the RMS radius of a
nucleus, i.e., with the volume of the nucleus where the nucleons
are confined. If we assume spherical symmetry, the mean volume of
the nucleus is
 \begin{equation}
 V=\frac{4\pi}{3}\langle r^2
\rangle^{3/2} = \frac{4\pi}{3} \left[4\pi \int_0^{\infty} r^4
\rho(r) d r\right]^{3/2}.
 \label{Volume}
\end{equation}
\begin{figure}[h]
\hspace*{-2cm}
\begin{center}
 {\includegraphics[width=5.5cm]{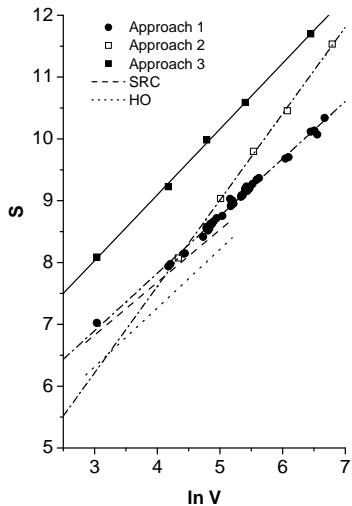}}
\end{center}
\vspace*{-1.2cm}
 \caption{\small The information entropies in nats for various nuclei versus
the mean volume of the nucleus. The lines correspond to the
fitting expression $S=a_V+b_V\ln V$. For the various cases see
text.}
\end{figure}

The calculated values of $S$ versus $\ln V$ for the various cases
we have examined are shown in Fig. 2. It is seen that the
information entropy sum depends linearly on the logarithm of $V$
 \begin{equation}
 S=a_V + b_V \ln V  .
 \label{S-V}
 \end{equation}
The values of the parameters $a_V$ and $b_V$ determined by least
square fit are given in Table I.

The almost parallel displacement of the lines corresponding to the
approaches 1 and 3 is due to the fact that we used the same DDs
from Refs. \cite{DeVries87,Burov98} in both approaches. The
different slope of the line of the second approach comes from the
CDFM we have used to calculate the DDs of the five nuclei from the
MD \cite{Antonov05}.

In Ref \cite{Hall99} it has been shown that for any ensemble
(classical or quantum, discrete or continous) there is essentially
only one measure of the "volume" occupied by the ensemble, which
is compatible with basic geometrical notions. This volume is
called "ensemble volume" and provides a universal choice or a
direct measure of uncertainty, which is advantageous when one
wishes to compare the spreads of two ensembles of a given type.
The ensemble volume turns out to be proportional to the exponent
of the entropy of the ensemble \cite{Hall99}, i.e.
\begin{equation}
V_{ensemble}=K(\Gamma)\ e^S .
 \label{V-ense}
 \end{equation}
The constant $K(\Gamma)$ is a normalization constant reflecting
the notion that only relative volumes are of real interest in
comparing different ensembles.

Using Eq. (\ref{V-ense}), the Gibbs relation $S_{therm}=kS$,
between thermodynamical entropy and ensemble entropy for
equilibrium ensembles, can be written as
\begin{equation}
 S_{therm}=k\ln[V_{ensemble}/K(\Gamma) ],
 \end{equation}
i.e., within an additive constant, the thermodynamical entropy is
proportional to the logarithm of the ensemble volume
\cite{Hall99}. We can make the same statement between the IE of a
nucleus and its mean volume, via Eq. (\ref{S-V}).

Assuming that Shannon's IE $S$ is the ensemble entropy of a
nucleus and substituting $S$ from Eq. (\ref{S-ab}) into Eq.
 (\ref{V-ense}), for $K(\Gamma)=1$, we have
 \begin{equation}
V_{ensemble}= c A^b\ , \quad c=e^a .
  \label{V-ense-A}
  \end{equation}
Thus, the ensemble volume of a nucleus is analogous to $A^b$.

It was mentioned in the introduction that Eq. (\ref{S-ab}) is an
universal property of the fermion systems (electrons in the atoms
or in metallic clusters, nucleons in nuclei) as well as of
correlated bosons in a trap
\cite{Gadre848587,Panos97,Massen9801,Massen02}. Thus, Eq.
(\ref{V-ense-A}) is also valid for atoms, metallic clusters and
for correlated bosonic systems.

From the linear dependence of $S$ on the logarithm of the mean
volume of the nucleons, Eq. (\ref{S-V}), and from Eq.
(\ref{V-ense}) a relation between the ensemble volume of the
nucleus and the mean volume of it can be found. That relation is
\begin{equation}
 V_{ensemble}= c_V V^{b_V}\ , \quad c_V= e^{a_V} .
 \label{V-ense-V}
 \end{equation}
Thus, the ensemble volume of a nucleus is analogous to $V^{b_V}$.

From the three approaches we have considered, the first and the
third are the most reliable ones because the DD's are based on
experimental data. In these two approaches the values of the
parameter $b_V$ are quite close to 1. That is why we could say
that the ensemble volume of a nucleus is analogous to the mean
volume of the nucleus.

\section{Conclusions}

A study of Shannon's IEs in coordinate space, $S_r$, in momentum
space, $S_k$, and their sum, $S$, was made for many nuclei using
three approaches based on "experimental" DDs or/and on
"experimental" MDs.

In the first approach we used Fermi-type phenomenological DDs
reproducing the electron scattering experiments for many nuclei
from $^4$He to $^{238}$U \cite{DeVries87,Burov98}. The MDs of
these nuclei were found within the CDFM
\cite{Antonov79,Antonov85,Antonov88,Antonov899486}. In the second
approach we used the "experimental" MDs from superscaling analysis
for the nuclei $^4$He, $^{12}$C, $^{27}$Al, $^{56}$Fe and
$^{197}$Au, \cite{Antonov05} while the DDs of these nuclei were
found within the CDFM. In the third approach we used the
"experimental" DDs of the five nuclei as in the first approach and
the "experimental" MDs as in the second approach. The DDs and the
MDs were used for the evaluation of the IEs $S_r$, $S_k$ and $S$.
It was found that in the three approaches $S_r$, $S_k$ and $S$
depend linearly on the logarithm of the mass number in accordance
with previous studies of various Fermi systems.

The values of $S$ found in the three approaches of the present
work were compared with the ones evaluated with the HO case and
the short-range correlated DDs and MDs. It was found that for all
the nuclei we have considered the following inequalities hold
\[
S_{HO} < S_{SRC} < S_{approach1} <  S_{approach2} < S_{approach3}
\]
Thus, the various cases can be classified according to either the
quantity of the experimental data we used or the values of the IE
sum obtained for the various nuclei. In other words the increase
of the quality of the DD or/and of the MD leads to an increase of
the values of the IE sum according the maximum entropy principle.

It is also found that $S_r$ depends strongly on $\ln A$ and does
not depend very much on the model we use. The behaviour of $S_k$
is opposite. The properties of $S_r$ and $S_k$ lead us to find
that, within an additive constant the information entropy $S$ is
proportional to the logarithm of the mean volume of the nucleus in
accordance with the fact that for equilibrium ensembles the
thermodynamical entropy is proportional to the ensemble
volume\cite{Hall99}. In the case that Shannon's entropy may be
considered as ensemble entropy, a connection of the ensemble
volume with the mass number of the nucleus or with the mean volume
of the nucleus can be found.

Finally, the comparison of the values of $S$ for the same nucleus
using phenomenological DDs from the analysis of different
experiments could be used to choose the most reliable one.

\section*{Acknowledgments}

The authors would like to thank Prof. H.V. Von Geramb and Dr. C.P.
Panos for their comments on the manuscript and for fruitful
discussions.

One of the authors (A.N.A.) is grateful to the Bulgarian National
foundation for the partial support under the Contract No.
$\Phi-$1416.

\end{document}